\documentclass[aps,prc,nofootinbib,amsmath,amssymb, showkeys,twocolumn]{revtex4}

\usepackage{graphicx}
\usepackage{dcolumn}
\usepackage{bm}
\usepackage{color}
\usepackage{mathrsfs}
\usepackage{comment}
\usepackage{mathtools}
\usepackage{float}
\usepackage{fancyhdr}
\begin{document}
\title{
Exploration of the solar system and beyond using a thermonuclear fusion drive}
\author{Roman Ya. Kezerashvili$^{1,2}$}
\affiliation{ \mbox{$^{1}$Physics Department, New York City College
of Technology, The City University of New York,} \\
Brooklyn, NY 11201, USA \\
\mbox{$^{2}$The Graduate School and University Center, The
City University of New York,} \\
New York, NY 10016, USA \\
rkezerashvili@citytech.cuny.edu}
\date{\today}

\begin{abstract}
It is demonstrated that the development of a nuclear fusion rocket
engine based on a D $-$ $^{3}$He (Deterium-Helium 3) technology will allow
to travel in the solar system and beyond with an ease never before attained
and opens new possibilities to humankind. The Direct Fusion Drive (DFD) is
the D $-$ $^{3}$He-fueled, aneutronic, thermonuclear
fusion propulsion system that is under development at Princeton University
Plasma Physics Laboratory \cite{Cohen2019}. It is shown that direct nuclear
fusion drive is even more important in human planetary exploration and
constitutes a solution to the problem of exploring the solar
system.

It is considered and analyzed the Earth -- Mars mission using the DFD. It is shown
that one-way trips to Mars in slightly more than 100 days become possible
and also journeys to the asteroid belt will take about 250 days \cite{GentaRK2020}.

It is presented an analysis of realistic new trajectories for a robotic mission
to Saturn's largest moon, Titan, to demonstrate the great
advantages related to the thermonuclear Direct Fusion Drive. The
trajectories calculations and analysis for Saturn's largest moon Titan
different profile missions are given based on the characteristics of a 2 MW
class DFD engine. This capability results in a total trip duration of 2.6
years for the thrust--coast--thrust profile and less than 2 years for the
continuous thrust profile \cite{GajeriRK2021}. Using the same 2 MW class DFD
engine one can reach some trans-Neptunian object, such as the dwarf planets
Makemake, Eris, and Haumea in less than 10 years with a payload mass of at
least 1500 kg, so that it would enable all kind of missions, from
scientific observation to in-situ operations \cite{AimeRK2021}. We
consider for each mission a Thrust-Coast-Thrust profile. For this reason,
each mission is divided into 3 phases: i. the spiral trajectory to escape
Earth gravity influence; ii. the interplanetary travel, from the exit of
Earth sphere of influence to the end of the coasting phase; iii. maneuvers
to rendezvous with Titan or the dwarf planet. We present calculations to reach a
vicinity at 125 AU for the study of the Sun magnetosphere.

Finally we conclude that a spacecraft propelled by DFD will open
unprecedented possibilities to explore the border of the solar system, in a
limited amount of time and with a very high payload to propellant masses
ratio.
\end{abstract}
\keywords{Direct Fusion Drive, Exploration of Mars, Exploration of Titan, Trans-Neptunian Objects Exploration, Deep Space Exploration}

\maketitle

\section{Introduction}

To develop a spacefaring civilization, humankind must develop spacecraft propulsion technologies that enable safe,
affordable and repeatable mobility through the solar system. The exploration of space started by using chemical propulsion rocket. Half a century ago human exploration of the solar system began with the closest celestial body and Earth's natural satellite, taking the first men to walk on lunar soil in 1969 \cite{Apollo 11}. Although human missions to Mars have not been attempted, it is possible
to forecast that the same technology is sufficient for preliminary missions \cite{1,2}. At the beginning of the
sixties, first robotic Mars missions were designed and launched \cite{Mars1,Mars2,Mars3} focusing on the mission time and payload capability. However, chemical propulsion system is faced with difficulties one of which is the maximum speed that a rocket can reach is limited by the rocket equation.

Future exploration of the solar system and beyond on a major scale requires propulsion systems
capable of performance much greater than is achievable with the present generation of rocket
engines using chemical propellants. It is essential to use nuclear energy
instead of chemical energy to propel spacecraft because of its inherent
superiority to chemical propulsion by seven orders of magnitude. There are two alternatives for nuclear propulsion system: nuclear fission and thermonuclear fusion propulsion systems. Both alternatives of Nuclear Thermal and Nuclear Electric
Propulsion (NTP and NEP) based on nuclear fission reactions have been
studied in detail, and the former was already bench-tested with very
satisfactory results. NTP and NEP can improve our chances of performing space missions by reducing the travel time and maximize the payload mass. However, fusion propulsion systems have
an order of magnitude more energy density and fusible elements are vastly more abundant in the cosmos than fissionable ones. As with NTP and NEP based on nuclear fission reactions, for fusion propulsion there are also two alternatives: electric and direct. The idea of using fusion power for spacecraft propulsion has
a long history \cite{Cooper1968,Angelo,Schulze1991,Bruno}, with its support arising from the high energy density of the fuel and the high velocity of the fusion products.
Early proponents of fusion rockets that provided steady propulsion based their designs on the fusion devices that are tokamaks \cite{Tokamak1,Tokamak2}, mirror machines \cite{Mirrormash} and levitated dipoles \cite{Dipole}. The experimental results of the early period in fusion history indicated that the plasma's anomalous transport, meaning poor plasma energy confinement, Deuterium-Tritium (D-T) burning, large and powerful machines, many meters in diameter, producing over a gigawatt in power and requiring a meter or more of neutron shielding. Such large and massive devices could not be launched fully assembled; upwards of 100 launch vehicles would be needed. Such daunting and expensive proposals never proceeded beyond the conceptual stage \cite{Cohen2019}. One of the approaches for fusion-based nuclear propulsion is the $Z$-pinch dense plasma focus method based on Magneto-Inertial Fusion that may potentially lead to a small, low cost fusion reactor assembly \cite{Lindemuth2009}. In early 1930s it was recognized that a pinch of sufficient strength, with the appropriate fuel, could conceivably be used to initiate fusion. A concept of $Z$-pinch fusion propulsion system was developed by NASA and private companies. The energy released by the Z-pinch effect accelerates lithium propellant to a high speed, resulting in a specific impulse value of 19436 s and thrust of 3.8 kN \cite{Miernik}. Project Icarus that started in 2009 is a theoretical design study for an unmanned, fusion-based probe to Alpha Centauri using current or near-future technology \cite{Kelvin2009,Freeland}. The leading design to come out of the project is a vessel dubbed "Firefly", built around a continuous, open-core $Z$-pinch D-T fusion engine based on experimental work at the University of Washington \cite{Shumlak}.\\
\indent New designs of fusion propulsion devices, where a fusion-reactor design is a field-reversed configuration, have raised optimism for the prospect of considerably smaller fusion-powered rockets that are far lighter, less radioactive, and less costly. Commensurate with their reduced size, these rocket engines would produce only megawatts of power \cite{Slough}, nevertheless ample for a wide variety of missions in the solar system and beyond. The common feature of these fusion propulsion devices is the geometry of the magnetic field that confines the plasma. We are considering a fusion-based reactor that utilizes D-$^3$He plasma known as the D-$^3$He-fueled Direct Fusion Drive (DFD) \cite{Cohen2019}.\\
\indent In this work, we present a review of a new class of missions in the solar system and beyond that will be enabled
by the realization of the DFD. The work is organized in the following way: in Section 2 is given
a brief description of the DFD engine, its main characteristics, fuel
choice, and the thrust model. Section 3 is devoted to considerations
of the missions design for Mars, Saturn's largest satellite Titan and trans-Neptunian planets, and 125 AU destinations, respectively. Finally, concluding remarks follow in Section 4.
\section{Direct Fusion Drive} \label{DFD}
The Direct Fusion Drive is a revolutionary fusion propulsion concept developed at Princeton Plasma Physics Laboratory (PPPL) \cite{cohen2011rf,Cohen2019} that would produce both propulsion and electric power from a single, compact fusion reactor \cite{razin2014direct}. The purpose of PPPL research project is to find solutions for the critical scientific and technological problems related to fusion technology. The DFD concept suits several kind of space destinations, such as Mars manned and robotic missions, heavy cargo missions to the outer solar system or the near interstellar space \cite{Cohen2019,thomas2017fusion, GentaRK2020,AimeRK2021,GajeriRK2021}.

The Princeton field reversed configuration (PFRC-2) concept employs a unique radio frequency (RF) plasma heating method, known as \textit{odd-parity heating},
which increases the plasma temperature in order to achieve the proper physics conditions, that enable the fusion process in a field reversed configuration (FRC) plasma \cite{cohen2000ion, cohen2007stochastic}. A new heating method, invented by Cohen and Milroy \cite{cohen2000maintaining}, is based on a magnetic field that is antisymmetric about the mid-plane normal to the axis and added to a FRC plasma maintaining its closed field line structure. It was first theorized in 2000 \cite{cohen2000maintaining} and demonstrated in 2006 (PFRC-1) \cite{cohen2007formation}. This is a crucial point because the open field lines let the plasma escape and consequently reduce confinement time, which is tightly bound to optimal fusion conditions \cite{guo2005observations}. The fusion process is magnetically confined in the core, the region inside the magnetic \textit{separatrix}, 
which is an imaginary closed surface that demarcates open magnetic field lines, those that cross the device walls from those that stay fully inside the device. The open field line region - also called the scrape-off layer (SOL) - is the region where the cold deuterium propellant is heated by the fusion products.

The PFRC-2 concept employs a unique radio frequency (RF) plasma heating method, known as \textit{odd-parity heating}, which increases the plasma temperature in order to achieve the proper physics conditions, which enable the fusion process in a field reversed configuration (FRC) plasma \cite{cohen2000ion, cohen2007stochastic}. The FRC is a particular magnetic field geometry, accidentally discovered in the sixties, in which a toroidal electric current is induced inside a cylindrical plasma, creating a poloidal magnetic field. The latter is reversed with respect to the direction of an externally applied axial magnetic field.

The fusion of light nuclei produces much more energy per unit mass than fission processes. Usually one of the components of the fusion reaction is proton, deuterium, or tritium. The other component involved into the fusion of light nuclei can be another deuterium, isotopes of helium, $_{2}^{3}$He or $ _{2}^{4}$He, and isotopes of lithium, $_{3}^{6}$Li and $_{7}^{7}$Li. The region, where fusion reactions take place in the DFD, is the high temperature, moderate density plasma region named the core. The fusion reaction of nuclei of deuterium ($_{1}^{2}$H), and tritium ($_{1}^{3}$H) is the most promising for the implementation of controlled thermonuclear fusion since its cross section even at low energies is sufficiently large \cite{dolan2013magnetic}. However, due to significant emission of neutrons, the D--T fuel is not the best choice for the DFD. In fact, an aneutronic fuel, such as the mixture of the deuterium and helium-3 ($_{2}^{3}$He) isotope, is most preferable. The choice of D--$_{2}^{3}$He fuel mixture to produce the D--$_{2}^{3}$He plasma is related to the neutrons production problem. If neutrons are produced from the fusion reactions, a certain amount of energy is not usable, leading to not negligible losses of energy, as well as the contamination of a spacecraft by the neutrons' emission, which should be shielded to protect the spacecraft and crew. Neutrons are hard to direct due to the fact that they have no charge and can not be controlled with electric or magnetic fields. Therefore, it is essential to reduce the neutron fluxes in order to minimize the damage and activation of nearby materials and structures. This would inevitably result in an increase of radiation shielding masses.

Let us focus on the fusion processes in D--$_{2}^{3}$He plasma. Depending on
plasma temperature the ignition of D--$_{2}^{3}$He, D$-$D, and $_{2}^{3}$%
He--$_{2}^{3}$He fusion reactions can occur in D--$_{2}^{3}$He plasma. Therefore, the D--$_{2}^{3}$He plasma can
admit the following aneutronic and neutron emitted primary reactions:
\begin{eqnarray}
\text{D}+\text{ }_{2}^{3}\text{He} &\rightarrow &\text{ }_{2}^{4}\text{He}+p%
\text{ \ \ \ \ \ \ \ }(Q=18.34\text{ MeV),}  \label{DD3He} \\
&\rightarrow &\text{ T}+2p  \label{DD3He1} \\
&\rightarrow &\text{ D}+\text{D}+p,  \label{DD3He2} \\
&\rightarrow &\text{ }_{3}^{5}\text{Li }+\gamma ;  \label{DD3He3} \\
\text{D}+\text{D} &\rightarrow &\text{ }_{2}^{3}\text{He}+n\text{ \ \ \ \ \
\ \ }(Q=3.25\text{ MeV}),  \label{DD} \\
&\rightarrow &\text{ T}+p\text{\ \ \ \ \ \ \ \ \ \ \ }(Q=4.04\text{ MeV);}
\label{DD1} \\
_{2}^{3}\text{He}+\text{ }_{2}^{3}\text{He} &\rightarrow &\text{ }_{2}^{4}%
\text{He}+2p\text{ \ \ \ \ \ \ }(Q=12.86\text{ MeV),}  \label{He3He3} \\
&\rightarrow &\text{ D}+\text{D}+2p,  \label{He3He31} \\
&\rightarrow &\text{ }_{3}^{5}\text{Li }+p,  \label{He3He32} \\
&\rightarrow &\text{ }_{3}^{5}\text{Li}^{\ast }\text{ }+p,  \label{He3He33}
\\
&\rightarrow &\text{ }_{4}^{6}\text{Be}+\gamma .  \label{He3He34}
\end{eqnarray}%
The $Q$ value for nuclear reactions (\ref{DD3He}), (\ref{DD})$-$(\ref{He3He3})  is the difference between the sum of the masses of the initial reactants and the sum of the masses of the final products. All primary reactions in D--$_{2}^{3}$He are aneutronic, with the exception of the process (\ref{DD}), where neutrons with the energy of 2.45 MeV are produced. The D--D has two almost equally exothermic channel (\ref{DD}) and (\ref{DD1}). The ignition of D--D fusion requires the plasma temperature about $5\cdot 10^8$ K. Due to production of tritium in D--D fusion (\ref{DD1}), D--$_{2}^{3}$He plasma admits the secondary fusion processes: deuterium--tritium, tritium--tritium,
and tritium--helium-3.
\begin{figure}[H]
\noindent
\begin{centering}
\includegraphics[width=7.5cm]{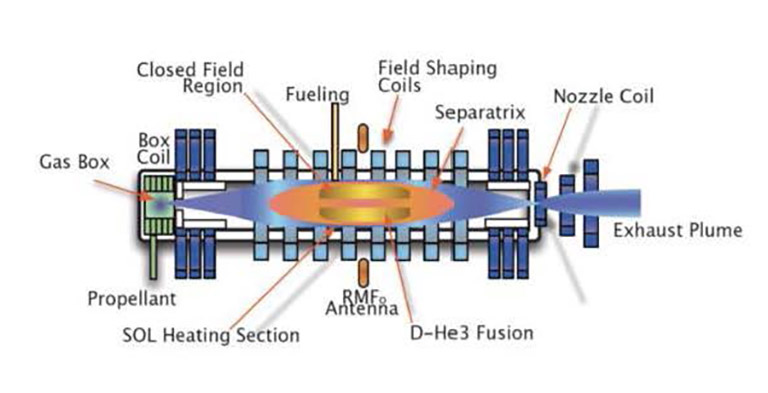}
\par\end{centering}
\caption{Schematic of a DFD with its simple linear configuration and
directed exhaust stream. A propellant is added
to the gas box. Fusion occurs in the closed-field-line region. Cool
plasma flows around the fusion region, absorbs energy from the fusion products, and is then accelerated by a magnetic
nozzle. Figure from Ref. \cite{Thomas2018}.}
\label{FigDFD1}
\end{figure}
Let us consider a D--$_{2}^{3}$He plasma. Our particular interest is for the process (\ref{DD}), where 2.45 MeV neutrons are produced. Also, reactions (\ref{DD3He1}) and (\ref{DD1}) are the
sources of tritium production, because they could happen as secondary reactions. In a hot D--$_{2}^{3}$He plasma the D--T and T--T fusion admits reactions where fast,
useless, and undesirable neutrons are emitted.
Undesired neutron contamination is among the main risk factors for the damaging of spacecraft materials.
Thus the problem of the contamination by the neutron emission in these processes exists if the produced tritium is not removed to prevent its fusion with deuterium, helium-3, and itself.

The $_{2}^{3}$He fuel consumption is very complex to calculate, and it depends on multiple factors. We base our estimate for fuel consumption on values calculated for other missions based on the DFD \cite{thomas2017fusion, paluszek2014direct}. The fuel consumption is calculated by dividing the total fuel consumption by the days of the mission and the fusion power considered for the previous studies \cite{thomas2017fusion, paluszek2014direct}. In this way, it has been found a fuel consumption per day per MW of power. This means that for a 2.5 years mission, under the hypothesis of a 2-MW class DFD engine, the mass of $_{2}^{3}$He would be about $0.27$ kg. This fuel mass value, on a spacecraft of multiple tons, can be neglected for all trajectories calculations.

The schematic of a DFD is shown in Fig. \ref{FigDFD1}. The heated plasma expands in a magnetic nozzle, converting its thermal energy into kinetic energy, thus providing thrust to the system. The magnetic nozzle  works as a physical nozzle, with the difference that the fluid does not directly hit any physical wall. Researchers at Princeton Satellite Systems (PSS) considered this data to produce a functional model of the thrust and specific impulse of the engine as a function of input power to the SOL and propellant flow rate. It is essential to underline that the input power into the SOL is only $40$-$50 \;\%$ of the total fusion power. If only fusion products were ejected directly from the engine, they would have a velocity of $25,000$ km/s producing negligible thrust \cite{thomas2017fusion}, but interacting with cool ionized gas in the SOL, the energy is transferred from the hot products to the electrons and then transferred to the ions as they traverse the magnetic nozzle. The resulting exit velocity is about $10^5$ m/s, generating a thrust of about $2.5$ to $5$ N per MW of fusion power \cite{thomas2017fusion,Cohen2019}, and from $5$ to $10$ N of thrust per MW of thrust power, with a specific impulse of about $10,000$ s. The estimated range for the specific power is between 0.75 kW/kg and 1.25 kW/kg. In Table \ref{DFDrange} the main characteristics for low and high power configurations of the engine are given. The DFD can be fully scaled in configuration and reach the required power. In our consideration we assume the minimum estimated value of 0.75 kW/kg of specific power, which is a conservative option for all the calculations and leads to an engine mass of 2660 kg.
\begin{table}[H]
\caption{Direct Fusion Drive performance. The characteristics for low and high power configuration are shown \cite{Cohen2019}.}
\centering
\begin{tabular}{r c c}
\hline \hline
     & Low power &  High power     \\
\hline
Fusion Power, MW        & 1  &      10       \\
Specific Impulse, s          & 8500 - 8000   &  12000 - 9900   \\
Thrust, N  &  4 \hspace{2mm} - \hspace{2mm} 5     &    35 \hspace{1mm} - \hspace{1mm} 55    \\
Thrust Power, MW          & 0.46    &      5.6     \\
Specific Power, kW/kg            &   0.75    &     1.25    \\
\hline \hline
\end{tabular}
\label{DFDrange}
\end{table}
\section{Missions design}
This section following Refs. \cite{GentaRK2020,AimeRK2021,GajeriRK2021} we focus on the missions design for Mars, Saturn's largest satellite Titan, trans-Neptunian
objects such as such as
the dwarf planets Makemake, Eris and Haumea, and 125 AU destinations, respectively, using the direct fusion drive. For these considerations
each mission is divided into following phases: i. the spiral trajectory to escape
the Earth gravity influence sphere; ii. the interplanetary travel, from the exit of
Earth gravity sphere of influence to the end of the coasting phase; iii. maneuvers
to rendezvous with Titan or trans-Neptunian dwarf planets. We present calculations to reach a
vicinity at 125 AU for the study of the Sun magnetosphere.
\subsection{Earth -- Mars Mission}
We studied the Earth -- Mars mission as well as Earth -- 16 Psyche asteroid mission (16 Psyche asteroid which belongs to the asteroid belt is about 1.1 AU far from the Mars) using DFD in detail and results are reported in Ref. \cite{GentaRK2020}. In our consideration of the Earth -- Mars mission we use the parameters for
the DFD thruster given in Table \ref{DFDrange}. As it was shown Ref. \cite{Genta}, a thruster with such a high specific
impulse and low specific mass must operate in a continuous thrust mode. A
study of an Earth-Mars transfer \cite{GentaRK2020}, was performed assuming that it can
operate in optimal Variable Ejection Velocity
conditions \cite{Genta} using the IRMA 7.1 computer
code \cite{22} with the following data: launch opportunity: 2037; specific
mass $\alpha =1.25$ kg/kW; overall efficiency $\eta =0.56$; tankage factor $%
k_{tank}=0.10$; height of circular starting Low Earth Orbit (LEO): 600 km; height of circular arrival Low Mars Orbit (LMO): 300 km.

\begin{figure}[H]
\noindent
\begin{centering}
\includegraphics[width=6.5cm]{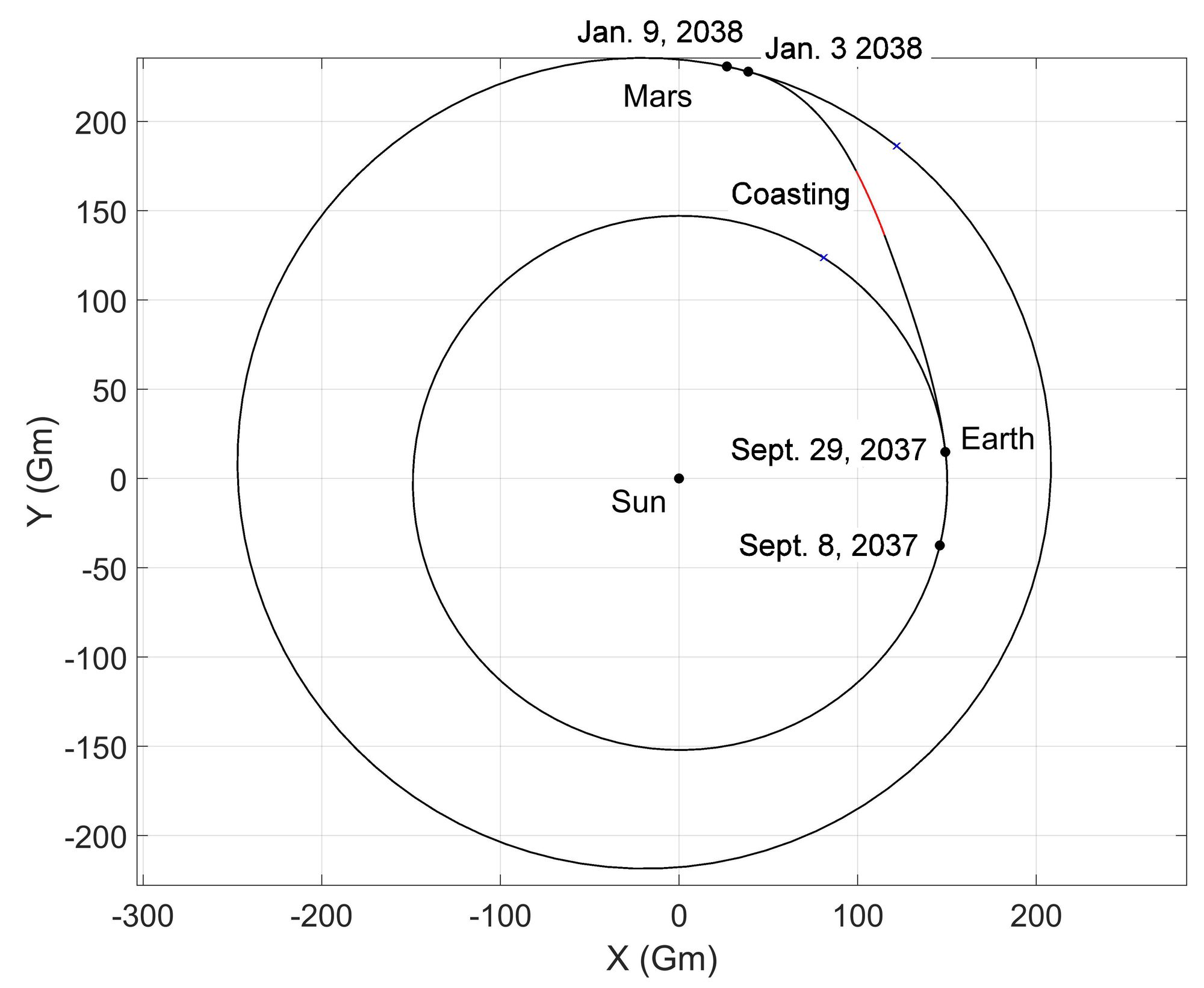}
\par\end{centering}
\caption{Earth-Mars trajectory for the
fast spacecraft. Figure from Ref. \cite{GentaRK2020}.}
\label{Fig3}
\end{figure}

\begin{figure}[H]
\noindent
\begin{centering}
\includegraphics[width=6.5cm]{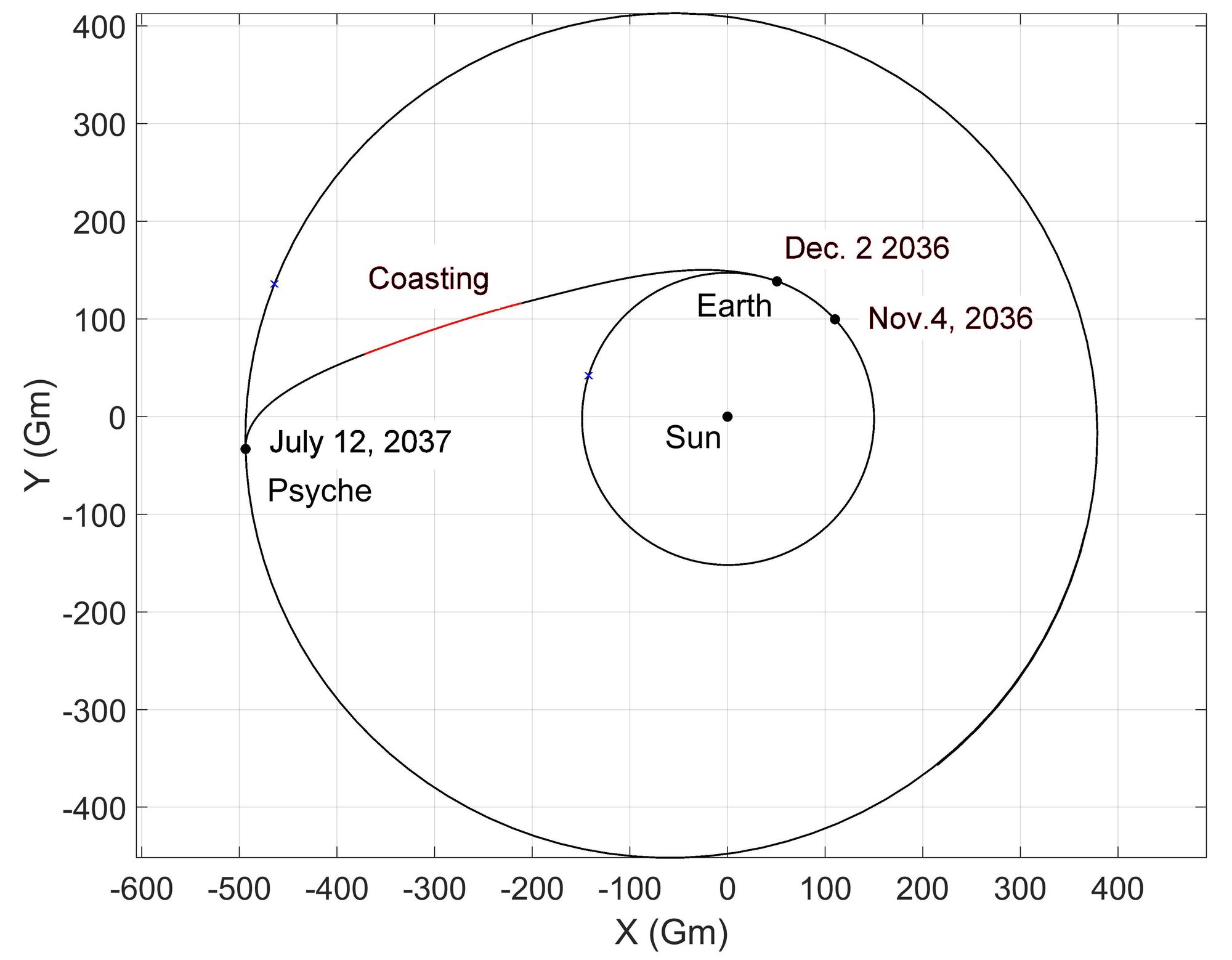}
\par\end{centering}
\caption{Trajectory for a 250 days
journey to the metal asteroid 16 Psyche. Figure from Ref. \cite{GentaRK2020}.}
\label{Fig5}
\end{figure}

The optimal trajectory for a 120 days Earth-Mars journey starts 66.6 days
before opposition, spends 8.4 days spiraling about Earth, 105.8 days in
interplanetary space, and finally 8.4 days spiraling about Mars to reach the
final LMO. The mass breakdown and the jet power are reported in Table \ref%
{Tab21} and the trajectory is shown in Fig. \ref{Fig3}.
\begin{table}[H]
\caption{Timing and mass breakdown of the Earth -- Mars and Earth -- 16 Psyche asteroid missions. Notations are the following: $t$ is time ($t_{d}$ -- days before opposition, $t_{t}$ -- days of the one-way trip, $t_{1}$ -- days at Low Earth Orbit, $t_{2}$ -- days in interplanetary space, $t_{3}$ -- days at Low Mars Orbit, respectively), $m_{i}$ is initial mass, $m_{p}$ is mass of payload, $m_{f}$ is mass of propellant, $m_{t}$ is mass of the thruster, mass, $m_{tank}$ is mass of tank, $P$ is  power of the jet.}
\label{Tab21}
\begin{center}
\begin{tabular}{lccc}
\hline\hline
Destination & \multicolumn{2}{c}{Mars} & 16 Psyche \\ \hline
Type & Optimal & Limited fast &  \\ \hline
$t_{d}$  & 66.6 & 71.9 & 120 \\
$t_{t}$  & 120 & 123 & 250 \\
$t_{1}$  & 8.4 & 20.6 & 27.6 \\
$t_{2}$  & 105.8 & 96.0 & 222.3 \\
$t_{3}$  & 5.8 & 6.4 & 0.1 \\
$(m_{l}+m_{s})/m_{i}$ & 0.254 & 0.258 & 0.241 \\
$m_{f}/m_{i}$ & 0.525 & 0.319 & 0.416 \\
$m_{p}/m_{i}$ & 0.169 & 0.391 & 0.169 \\
$m_{tank}/m_{i}$ & 0.052 & 0.0319 & 0.042 \\
$P/m_{i}$ (W/kg) & 75.62 & 175.35 & 134.60 \\ \hline\hline
\end{tabular}
\end{center}
\end{table}
Asteroid 16 Psyche is one of the most massive of the metal-rich M-type asteroids extremely rich in nickel and iron, but also in gold and NASA plans a
mission to survey this asteroid in August of 2022,
and arrive at the asteroid in early 2026, following a Mars gravity assist in
2023. A mission to the same asteroid using the DFD thruster can be performed in a quite short time: for instance, using the launch
opportunity of 2037 (the opposition is on March 4, 2037) and starting 120
days before the opposition, a mission lasting only 250 days can be
performed. This figure must be compared with the roughly 3.5 years of the
mentioned NASA proposal, based on chemical propulsion and gravity assist.
The mass breakdown and the jet power for Earth -- 16 Psyche asteroid mission are reported in Table \ref{Tab21} and the trajectory is shown in Fig. \ref{Fig5}.
\subsection{Earth -- Titan mission}
\begin{figure}[H]
  \centering
  \includegraphics[width=7.5cm ]{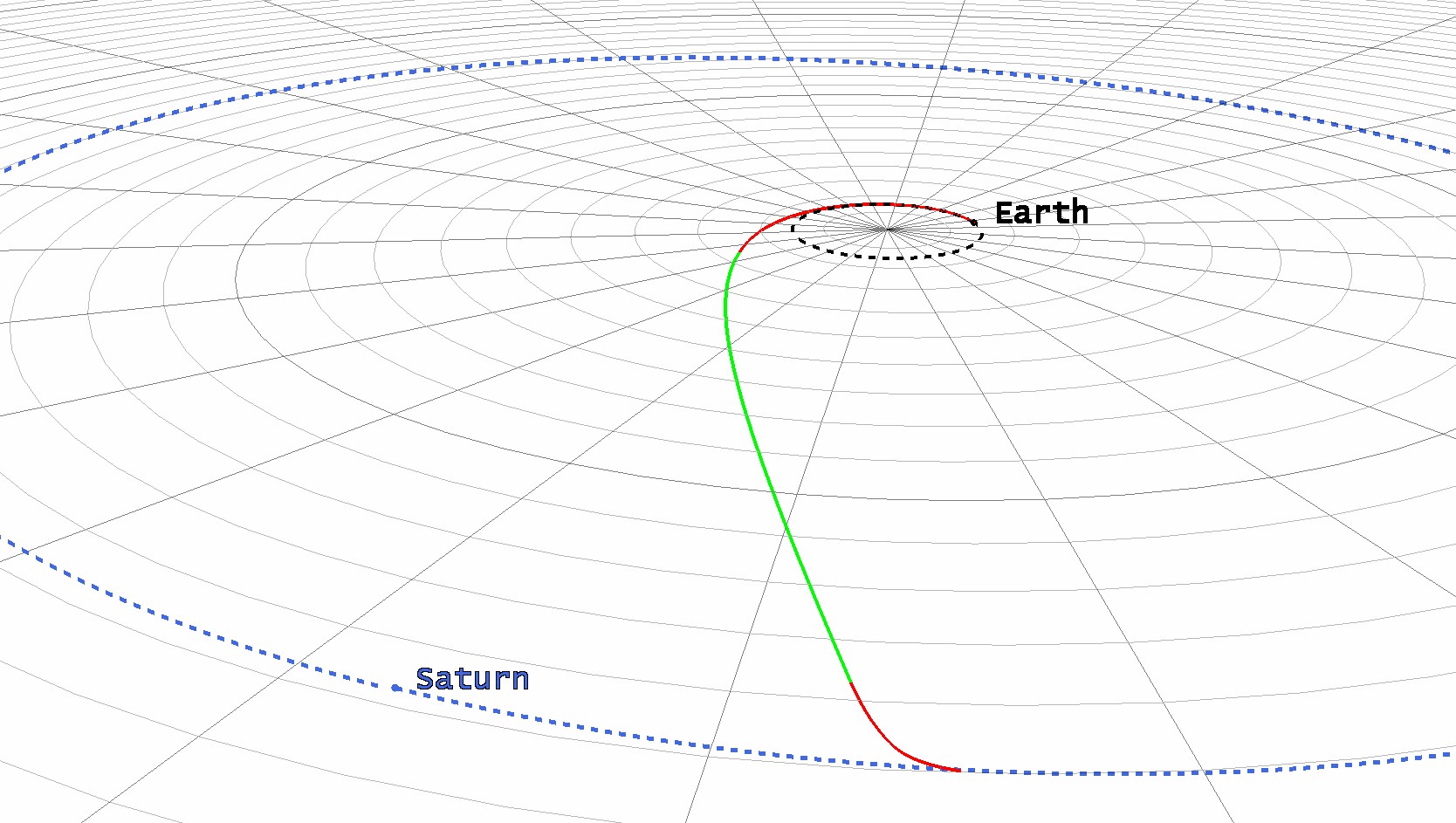}
  \caption{Thrust-Coast-Thrust profile for the Titan mission. It is possible to observe three segments of the trajectory, the red solid curves indicate that the spacecraft thrust is active, and the green line represents the coasting phase without active thrust. Figure from Ref. \cite{GajeriRK2021}.}
  \label{fig:tct}
\end{figure}

\begin{figure}[H]
  \centering
  \includegraphics[width=7.5 cm]{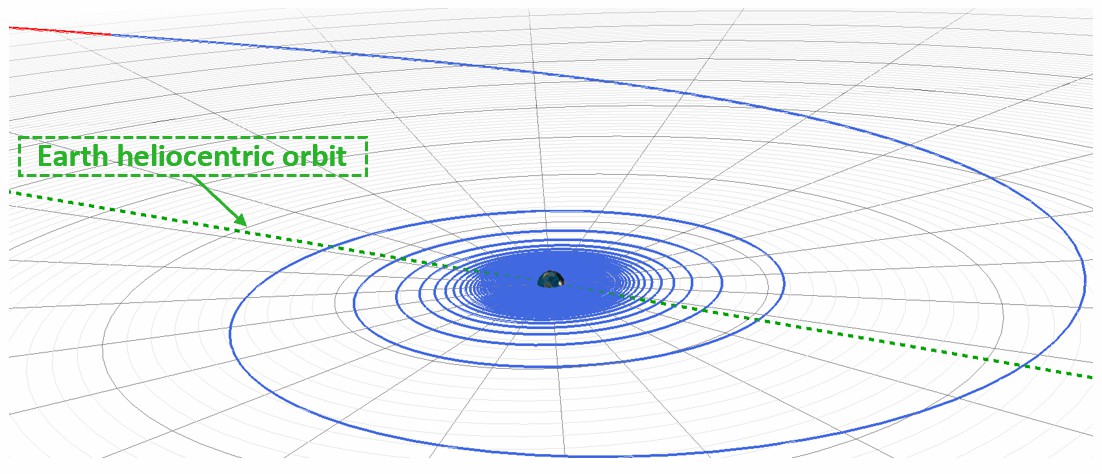}
  \caption{Spiral trajectory for the escape maneuver from the Earth gravitational sphere of influence. The escape trajectory (blue) is shown using the Earth centered reference system. The red solid curve represents the spacecraft trajectory, once it is outside the Earth's gravitational sphere of influence, moving into interplanetary space. Figure from Ref. \cite{GajeriRK2021}.}
  \label{fig:Escape}
\end{figure}

\begin{figure}[H]
\centering
\includegraphics[width=7.5 cm]{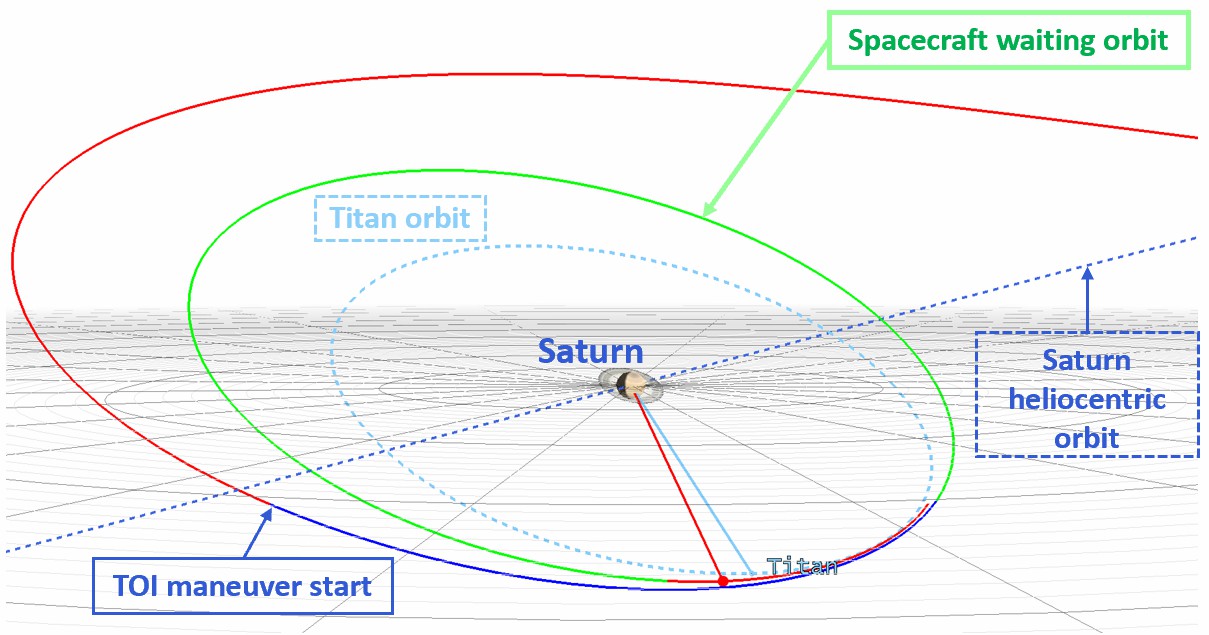}
\caption{Titan orbit insertion phase. Central body: Saturn. The blue line marks the start of the required TOI maneuver, subsequent the SOI maneuver (red). It is possible to observe the portion of the waiting orbit (green) traveled by the spacecraft before the phasing maneuver starts (red). Figure from Ref. \cite{GajeriRK2021}.}
\label{fig:phasing_1}
\end{figure}

\begin{figure}[H]
  \centering
  \includegraphics[width=7.5cm]{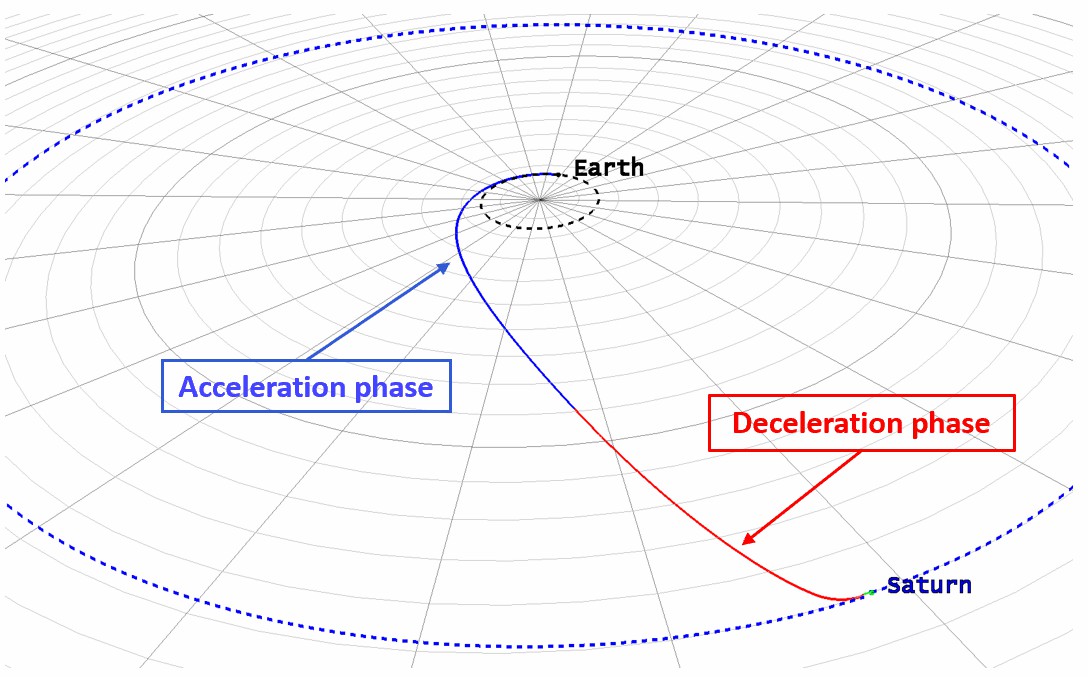}
  \caption{Planar trajectory for the continuous thrust profile mission. At the end of the blue curve there is the change in direction of the thrust (switch time). The trajectory follows Earth's orbit for some time before a nearly straight trajectory to Saturn. Figure from Ref. \cite{GajeriRK2021}.}
  \label{fig:profiloTcontinua}
\end{figure}

The objective of the mission is to reach Saturn near the descending node referred to the ecliptic plane along its orbit around the Sun and solve a nearly 2D problem, with huge advantages at numerical and computational levels. Once the spacecraft is orbiting around Saturn, the Titan orbit insertion maneuver concludes the mission. In Ref. \cite{GajeriRK2021} we considered two profiles for the Earth -- Titan mission: Thrust-Coast-Thrust (T-C-T) and Continuous Thrust (CT) profiles. The T-C-T profile mission is divided into four different phases: Earth departure, interplanetary trajectory, Saturn orbit insertion, and Titan orbit insertion. The mission start time has been estimated by considering both the Earth and Saturn orbits and by taking into account the time constraint represented by the rendezvous with the target planet. The inputs for this rendezvous problem are the DFD engine parameters, initial and radii for departure and arrival orbits, and the initial guesses for the payload and spacecraft masses. Travel time and mission start time estimations are obtained after several iterations, starting from initial guess related to the impulsive approach and considering some crucial constraints, for instance, related to the total mass of the spacecraft. In Fig. \ref{fig:tct} the first three phases of the mission are shown. The first red solid curve starts from the Earth initial position and contains both the escape maneuver from Earth and the burn which puts the spacecraft along the heliocentric hyperbolic trajectory to Saturn. The second red solid curve shows the Saturn orbit insertion. While the green curve indicates the interplanetary trajectory without thrust.
The Earth departure phase is to escape the Earth's gravitational sphere of influence and insert the spacecraft directly into a heliocentric orbit. Our simulations show
that the Earth departure from a LEO uses reasonable
propellant mass and takes between 25 and 71 days depending on the DFD engine parameters and initial mass considered. The initial orbit is a circular orbit with an
altitude of about 386 km and inclination of about 24 degree which
allows the spacecraft to escape from Earth gravitational influence along
the Ecliptic plane. A spiral trajectory for the escape maneuver from the Earth gravitational sphere of influence is shown in Fig. \ref{fig:Escape}.
Once the Earth departure phase is completed, and the spacecraft is in an elliptic orbit around the Sun, the interplanetary trajectory mission phase begins. The main goal of this phase is to reach a spatial region on the ecliptic plane, pointing at the descending node of Saturn's heliocentric orbit. Many days of acceleration are needed at this point to obtain a heliocentric hyperbolic orbit, where the thrust vector is aligned with the velocity of the spacecraft. It is necessary to include a 1.6 year long coasting segment before the deceleration maneuver when the spacecraft is approaching the target planet, ensuring that the spacecraft and Saturn velocities are comparable. The Saturn orbit insertion (SOI) maneuver puts the spacecraft into Saturn's orbit.
The main purpose of the SOI is to obtain a proper velocity vector that allows the spacecraft to orbit around Saturn at a radius comparable to that of Titan. Let us adopt a Saturn centred reference system. It is required that after the SOI maneuver the spacecraft orbit switches from a hyperbolic trajectory to an elliptical orbit around the target planet. Different solutions are considered for the Titan orbit insertion (TOI) phase. In order to properly deal with the mutual gravitational interaction between Saturn, Titan, and the spacecraft, it is necessary to address the three-body problem. This problem consists of determining the perturbations in the motion of one of the bodies around the central body, produced by the attraction of the third. A strong time constraint has to be considered to solve this problem. In Ref. \cite{GajeriRK2021} we adopted the following maneuver strategy: i) a  first finite burn where the thrust vector is directed along the anti-velocity direction, ii) a coasting segment, iii) a phasing maneuver where thrust components along anti-velocity
and co-normal direction are considered, iv) the final Titan orbit insertion maneuver.
\noindent Once the spacecraft reaches Titan with the proper relative velocity, the final orbit insertion maneuver starts, achieving a Titan centered circular orbit 4000 km away from the surface, as shown in Fig. \ref{fig:phasing_1}. This altitude provides sufficient orbital stability, by requiring less station keeping maneuvers to fight Saturn's perturbations, which can be useful to freely vary the spacecraft orbit around Titan. Moreover, in order to meet the scientific instruments requirements, it is possible to achieve closer distance to the surface by varying thrust direction and duration with a negligible amount of propellant.
The natural alternative solution to the T-C-T mission profile is represented by the CT profile mission where the engine is always on, generating a constant thrust during the previous four different mission phases. The main goal of the CT scenario calculation is to find a proper switch time for the thrust direction inversion in order to reach Saturn with an acceptable velocity, allowing the spacecraft to orbit the planet. The same scenario as for the T-C-T mission is used to the Saturn and Titan orbit insertion maneuvers. It is worth noting that in order to achieve a comparable time duration of the escape maneuver with respect to the T-C-T solution, a decrease of payload
mass is essential for the CT mission. The trajectory for the CT profile mission is shown in Fig. \ref{fig:profiloTcontinua}.

Let us consider the T-C-T and CT missions and compare them to one of the most successful scientific mission, Cassini-Huygens, which studied the Saturn system for more than 10 years. A summary for the entire studied scenarios are given in Table \ref{comparison1vera}, in order to make a comparison of mission durations, payload masses, and propellant consumptions. In Table \ref{comparison1vera} along with the results of our calculations are also presented data for the Cassini-Huygens mission taken from Refs. \cite{mitchell2006cassini, NASA, ESA, johnson2005power}.
\begin{table*} [ht]
\caption{Comparison between the thrust-cost-thrust profile, the continuous thrust profile, and Cassini-Huygens missions. The CT profile mission results into an even shorter time travel - less than two years - still with a heavier payload than previous missions.}
\centering
\begin{tabular}{l c c c }
\hline \hline
		& T-C-T profile & CT profile & Cassini-Huygens mission \\
\hline
Travel time (to Saturn), [days] &  958.50 
&  714.05  & 2422 \\ 
Initial spacecraft mass, [kg] & 7250 &  9015 & 5712 \\
Payload mass, [kg] & 1800  &  1000 & 617.4 \\
Propellant mass used, [kg]  & 2658 &  5347 & 2950 \\ 
Fuel ($_{2}^{3}$He) mass used, [kg]  & 0.282 & 0.201 & --\\
Maximum trip velocity (Sun), [km/s] & 34.560 & 47.814 & --\\
\hline \hline
\end{tabular}
\label{comparison1vera}
\end{table*}
It is worth noticing that in the CT case, the DFD is capable of really fast travels, rapidly reaching extremely high velocity. In the last mission phases, when a great amount of propellant has already been consumed and the spacecraft mass is decreased, the DFD can reach the required speed in a relatively short period. As a result, for the CT profile mission, the payload mass would be decreased because of the higher initial propellant mass, with the advantage of reducing the time of flight by over 250 days. The reduction in payload mass is mainly due to the Earth departure phase: the higher the initial mass, the longer the time necessary to escape from Earth. The payload mass of about 1000 kg is obtained through an iterative process, in order to provide a relatively fast spiral Earth departure comparable with the T-C-T scenario solution. Collectively, such kinds of maneuvers would be too demanding for any kind of present propulsion systems. The total trip duration is below 2 years for the CT profile mission, which is more than three times less the duration of the Cassini spacecraft travel to Saturn, which has been possible due to several gravity assists. It is important to emphasize that the payload mass has increased significantly, delivering 1000 kg in the fastest solution or 1800 kg in the T-C-T profile mission. For comparison, the Cassini spacecraft had a total payload of about 617 kg, including the Huygens lander (349 kg) \cite{harland2002mission}. Another advantage related to the shortening of total mission duration is the reduction of precious fuel mass (Helium 3), compared with the scenario of the T-C-T profile mission.

There are two possible solutions related to the operative phase of the mission. The results lead to two different feasible mission concepts. The high payload capability for both mission profiles allows to consider a parachute descent through Titan's dense atmosphere performed by a lander probe, containing a rover or even better a rotorcraft, carried on the main spacecraft (orbiter) \cite{tobie2006episodic,elachi2005cassini}. In this case, during the TOI maneuver, the orbiter releases the lander and keeps orbiting around Titan or, by performing a proper maneuver, can orbit again around Saturn.

In summary, the T-C-T mission profile is based on the assumption that the DFD will be capable of turning off and on the thrust generation. This is an important hypothesis which requires that the engine will not produce thrust during the coasting phase, which is in theory possible but not yet certain. More specifically, because of the robotic nature of the mission, it could be possible to think of turning off the engine in order to save both deuterium propellant and precious $_{2}^{3}$He fuel. The total fuel consumption for the entire mission is $\approx 0.112$ kg. Another possible solution, which could be more feasible, is based on the DFD ability to turn off and on the thrust generation without shutting down the engine, still generating the electrical power from the nuclear fusion reactions. In this case, the reactor still provides energy for the entire mission, and the $_{2}^{3}$He consumption would be around $0.282$ kg.
\subsection{Exploration of trans-Neptunian objects}
\begin{figure} [t]
\centering
\includegraphics[width=8.5cm]{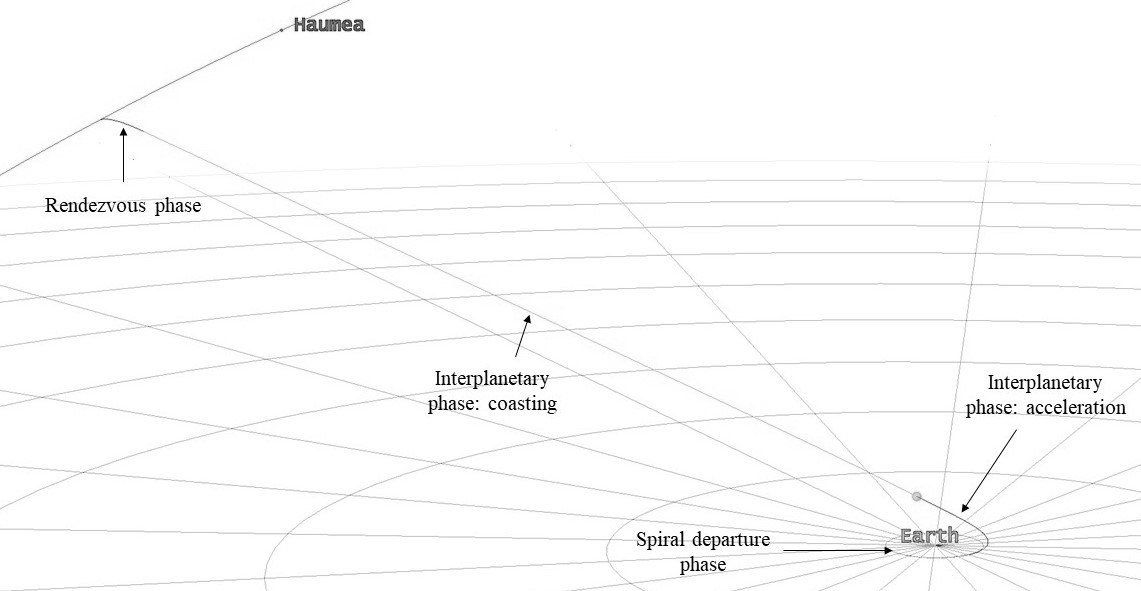}
\caption{The trajectory to reach Haumea. Each phase as well as the Earth orbit and a section of Haumea orbit in the bottom right and top left, respectively, are shown in this figure. The grid represents the ecliptic plane. Figure from Ref. \cite{AimeRK2021}. }
\label{fig:traj}
\end{figure}

The objective is to reach some trans-Neptunian object, such as
the dwarf planets Makemake, Eris, and Haumea in less than 10 years with a payload mass of at least of 1500 kg, so that it would enable all kinds  of missions, from
scientific observation to in-situ operations. In Ref. \cite{AimeRK2021} is performed the detailed study of the missions for the exploration trans-Neptunian dwarf planets and for 125 AU destinations using the nuclear fusion-based DFD.   For trajectories design for exploration of trans-Neptunian objects it is considered a 2 MW DFD and assume that thrust
and specific impulse are constant. The power transferred to the SOL is
supposed of about 1 MW, with the resulting performance of 8 N of
thrust and approximately 10$^{4}$ s of specific impulse.
The trajectory profile chosen is the Thrust-Coast-Thrust profile. Missions three phases are the following: a spiral departure phase from the LEO with the exit
from the sphere of influence of the Earth, an interplanetary phase, and a rendezvous phase wish utilizes of all the
maneuvers necessary to rendezvous with the dwarf trans-Neptunian planet destination. The inputs for all calculations are the engine performance and the
launch mass. The objective is to bring at least 1500 kg of payload at
destination in less than 10 years. Therefore, of a specific power of 1
kW/kg DFD results in approximately 2000 kg of engine, thereby a total dry
mass of 3500 kg. At this point, the first estimation of propellant needed
is 3988 kg, for a total launch mass of 7488 kg.\\
\indent For each mission the three phases shown in Fig. \ref{fig:traj} are studied. Results for the first phase are the same for all the missions. The only difference between the three missions is the departure date. This date is chosen to be later than the year 2050 because the three planets are moving towards the line of nodes, and the later will be the departure, the less inclination change will be required. Though, this is not a big restriction, because the difference is only of some days of acceleration, and for this reason it is negligible.
\begin{table*} [ht]
\caption{Summary for the three real case missions, with focus on the interplanetary and rendezvous phase maneuvers. It is worth noticing that the three sets of results are for scenarios with the same initial mass of 7488 kg. The final velocity for the interplanetary phase is the velocity at the end of the acceleration. The final coasting velocity is the one after the gravity losses, before rendezvous maneuvers. The final rendezvous velocity is the one at the end of the rendezvous phase. In the mission summary the time of flight is the total flight duration from the departure from LEO, the inclination is referred to the dwarf planet with respect to the ecliptic plane, the final mass delivered is the payload mass plus both the engine mass and the propellant mass not used. The $\Delta V$ in the mission summary is the sum of that of all thrust phases, including the spiral phase.} 
\centering
\begin{tabular}{ccccc}
\hline\hline
&  & Haumea & Makemake & Eris \\ \hline
Interplanetary & Acceleration duration, days & 225 & 251 & 279 \\
phase & Propellant consumption, kg & 1586 & 1769 & 1966 \\
& $\Delta V$, km/s & 25.47 & 28.89 & 32.71 \\
& Final velocity, km/s
 & 38.22 & 37.86 & 43.33 \\
& AU/year & 8.05 & 7.98 & 9.13 \\
& Coasting duration, years & 4.75 & 5.83 & 8.67 \\ \hline
Rendezvous & Maneuver duration, days & 206 & 204 & 243 \\
phase & Propellant consumption, kg & 1455 & 1436 & 1709 \\
& $\Delta V$, km/s & 31.15 & 31.96 & 41.37 \\
& Coasting velocity, km/s & 30.87 & 30.99 & 38.52 \\
& Velocity at rendezvous, km/s & 5.35 & 4.58 & 3.09 \\ \hline
Mission & Time of flight, years & 6.08 & 7.25 & 10.33 \\
summary & Distance at rendezvous, AU & 36.51 & 44.35 & 78.20 \\
& Inclination, deg & 28.19$^{0}$ & 29.01$^{0}$ & 43.87$^{0}$ \\
& Final mass delivered, kg & 3892 & 3733 & 3257 \\
& Propellant consumption, kg & 3595 & 3754 & 4231 \\
& $\Delta V$, km/s & 64.16 & 68.40 & 81.63 \\ \hline\hline
\end{tabular}
\label{tab:maneuverscomp}
\end{table*}
In Table \ref{tab:maneuverscomp} the summary for the interplanetary and rendezvous maneuver phases for the three real case missions is presented. $\Delta V$ differences and propellant consumption from Table \ref{tab:maneuverscomp} are mainly linked to the distance of the destination to reach and the final inclination to achieve, as well as the overall flight time. This means that, in order to keep the flight time reasonably low, the acceleration phase lasts a little longer for missions to the furthest dwarf planets. Also, part of the total acceleration duration is due to the required change of inclination: higher elevation from the ecliptic plane requires an increase in the out of plane thrust component direction, but this limits the possible acceleration, thereby there is a need to increase the thrust days. A trade-off between thrust vector direction and thrust days has been performed to minimize the fuel consumption. As far as the rendezvous is concerned, for the Eris mission the $\Delta V$ is about 10 km/s more demanding: this is because, in addition to the higher final coasting velocity, there is the fact that the velocity vector for that case has to be turned of about 120$^{0}$ to match Eris velocity, while for the other two cases the angle between the initial and final velocities is about 90$^{0}$.
\subsection{Exploration of Sun magnetosphere}
The primary objective of this mission is to reach 125 AU to study the magnetosphere of the Sun, but in its voyage, it will also flyby Eris dwarf planet. Eris is chosen as the flyby destination because it is the most representative dwarf planet, due to its highly inclined and eccentric orbit and to the planet's physical characteristics, but any other trans-Neptunian  dwarf planet can be considered for the flyby.
Results of calculations for the acceleration and coasting phases for the 125 AU mission with Eris flyby are given in Table \ref{tab:125eris}. As it is seen from Table \ref{tab:125eris}, in a little less than 9 years this mission aims to reach enormous distances from the Sun, visiting Eris and then fly towards interstellar space. This is possible because of the tremendous velocity reached at the end of the acceleration phase, which remains quite unaffected throughout the entire flight.
\begin{table} [H]
\centering
\caption{Acceleration and coasting phase for the 125 AU mission with Eris flyby. The initial Sun distance is the distance at which the thrust is turned off, while the distance at flyby is the distance of Eris from the Sun at the moment of flyby.}
\begin{tabular}{ r c }
\hline \hline
 & Acceleration phase \\
\hline
Duration & 510 days \\
$\Delta V$ & 71.50 km/s \\
Fuel used & 3595 kg \\
Final velocity & 77.41 km/s (16.33 AU/yr)\\
\hline
 & Coasting phase \\
\hline
Duration & 7 years\\
Initial Sun distance & 11.34 AU \\
Distance at flyby & 80.23 AU\\
Final Sun distance & 125 AU \\
Velocity at flyby & 76.55 km/s \\
\hline
 & Overall \\
\hline
Total mission duration,  & 8.67 years \\
Total $\Delta V$ & 79.05 km/s \\
Final mass delivered & 3350 kg \\
Propellant mass & 4150 kg \\
\hline \hline
\end{tabular}
\label{tab:125eris}
\end{table}
\section{Concluding remarks}
In this work, it is demonstrated that a spacecraft propelled by the
DFD will enable an entirely new class of missions and will pave the
way to a first practical approach to interstellar travel.
From the studies presented here, it is clear that the development of a
nuclear fusion rocket engine based on the D-$^{3}$He technology will allow
traveling in the solar system with an ease never before attained, opening
great possibilities to humankind.
One way trips to Mars in slightly more than 100 days become
possible and also missions to the asteroid belt in about 250 days.

The spacecraft with DFD based on aneutronic D-$^{3}$He still requires much
research and development, but it is possible that they become feasible
in less than two or three decades. The launch opportunity here studied
is that of 2037, 16 years from now which is a fairly favorable one for
Mars, while, on the contrary is not a very good one for 16 Psyche.
Our analysis confirm that a Titan mission using a 2-MW class DFD engine is not only feasible, but the departure
from LEO dramatically reduces launch and overall mission costs. We demonstrate
that the total Earth -- Titan mission duration is about 2.6 years for
the T-C-T profile, and less than 2 years for the CT profile, which is
more than three times less the duration of the Cassini spacecraft travel
to Saturn.
One of our main objectives is to present
the possibilities to visit and even rendezvous with destinations such as
TNOs and beyond using the DFD, given the time of flight and the launch
mass. The strong advantages related to this new propulsion technology result
in a great reduction of travel time with respect to the previous performed
missions and a tremendous payload increase with a huge availability of on-board electrical power.
The DFD would be a true game-changer for any robotic missions to asteroids, solar system
planets, and moons. Any other deep space mission becomes faster and
cheaper.\\
\indent Today chemical propulsion technologies are available to make a
Mars mission possible and the foundation of fusion propulsion is already
being built. However, it could be a nuclear fusion-powered spacecraft
that ferries us to Mars in foreseeable future. We should believe that by
mid-21st century, trips to Mars may become as routine as trips to the
International Space Station today due to achievements in developments
of fusion-powered thrusters \cite{GentaRK2020}.

\acknowledgments

The author is grateful to G. Genta, P. Aime, M. Gajeri for the fruitful collaborations that resulted in the publications \cite{GentaRK2020}, \cite{AimeRK2021}, and \cite{GajeriRK2021}, and thanks A. Spiridonova for helpful discussions.

\end{document}